\pdfoutput=1
\documentclass[12pt,a4paper,titlepage]{article}
\usepackage[utf8]{inputenc}
\usepackage[top=50pt,bottom=50pt,left=68pt,right=66pt]{geometry}
\usepackage{amsmath}
\usepackage{booktabs}
\usepackage{amssymb}
\usepackage[title]{appendix}
\usepackage{mathrsfs}
\usepackage{float}
\usepackage{multirow}
\usepackage{graphicx,caption,subcaption}
\usepackage[space]{grffile}

\begin{document}
\renewcommand{\arraystretch}{1.3}

\makeatletter
\def\@hangfrom#1{\setbox\@tempboxa\hbox{{#1}}%
      \hangindent 0pt
      \noindent\box\@tempboxa}
\makeatother


\def\un#1{\relax\ifmmode\@@underline#1\else
        $\@@underline{\hbox{#1}}$\relax\fi}


\let\under=\unt                 
\let\ced=\ce                    
\let\du=\du                     
\let\um=\Hu                     
\let\sll=\lp                    
\let\Sll=\Lp                    
\let\slo=\os                    
\let\Slo=\Os                    
\let\tie=\ta                    
\let\br=\ub                     


\def\a{\alpha}
\def\b{\beta}
\def\c{\chi}
\def\d{\delta}
\def\e{\epsilon}
\def\f{\phi}
\def\g{\gamma}
\def\h{\eta}
\def\i{\iota}
\def\j{\psi}
\def\k{\kappa}
\def\l{\lambda}
\def\m{\mu}
\def\n{\nu}
\def\o{\omega}
\def\p{\pi}
\def\q{\theta}
\def\r{\rho}
\def\s{\sigma}
\def\t{\tau}
\def\u{\upsilon}
\def\x{\xi}
\def\z{\zeta}
\def\D{\Delta}
\def\F{\Phi}
\def\G{\Gamma}
\def\J{\Psi}
\def\L{\Lambda}
\def\O{\Omega}
\def\P{\Pi}
\def\Q{\Theta}
\def\S{\Sigma}
\def\U{\Upsilon}
\def\X{\Xi}


\def\ve{\varepsilon}
\def\vf{\varphi}
\def\vr{\varrho}
\def\vs{\varsigma}
\def\vq{\vartheta}


\def\ca{{\cal A}}
\def\cb{{\cal B}}
\def\cc{{\cal C}}
\def\cd{{\cal D}}
\def\ce{{\cal E}}
\def\cf{{\cal F}}
\def\cg{{\cal G}}
\def\ch{{\cal H}}
\def\ci{{\cal I}}
\def\cj{{\cal J}}
\def\ck{{\cal K}}
\def\cl{{\cal L}}
\def\cm{{\cal M}}
\def\cn{{\cal N}}
\def\co{{\cal O}}
\def\cp{{\cal P}}
\def\cq{{\cal Q}}
\def\car{{\cal R}}
\def\cs{{\cal S}}
\def\ct{{\cal T}}
\def\cu{{\cal U}}
\def\cv{{\cal V}}
\def\cw{{\cal W}}
\def\cx{{\cal X}}
\def\cy{{\cal Y}}
\def\cz{{\cal Z}}


\def\Sc#1{{\hbox{\sc #1}}}      
\def\Sf#1{{\hbox{\sf #1}}}      



\def\slpa{\slash{\pa}}                            
\def\slin{\SLLash{\in}}                                   
\def\bo{{\raise-.3ex\hbox{\large$\Box$}}}               
\def\cbo{\Sc [}                                         
\def\pa{\partial}                                       
\def\de{\nabla}                                         
\def\dell{\bigtriangledown}                             
\def\su{\sum}                                           
\def\pr{\prod}                                          
\def\iff{\leftrightarrow}                               
\def\conj{{\hbox{\large *}}}                            
\def\ltap{\raisebox{-.4ex}{\rlap{$\sim$}} \raisebox{.4ex}{$<$}}   
\def\gtap{\raisebox{-.4ex}{\rlap{$\sim$}} \raisebox{.4ex}{$>$}}   
\def\TH{{\raise.2ex\hbox{$\displaystyle \bigodot$}\mskip-4.7mu \llap H \;}}
\def\face{{\raise.2ex\hbox{$\displaystyle \bigodot$}\mskip-2.2mu \llap {$\ddot
        \smile$}}}                                      
\def\dg{\sp\dagger}                                     
\def\ddg{\sp\ddagger}                                   

\font\tenex=cmex10 scaled 1200


\def\sp#1{{}^{#1}}                              
\def\sb#1{{}_{#1}}                              
\def\oldsl#1{\rlap/#1}                          
\def\slash#1{\rlap{\hbox{$\mskip 1 mu /$}}#1}      
\def\Slash#1{\rlap{\hbox{$\mskip 3 mu /$}}#1}      
\def\SLash#1{\rlap{\hbox{$\mskip 4.5 mu /$}}#1}    
\def\SLLash#1{\rlap{\hbox{$\mskip 6 mu /$}}#1}      
\def\PMMM#1{\rlap{\hbox{$\mskip 2 mu | $}}#1}   %
\def\PMM#1{\rlap{\hbox{$\mskip 4 mu ~ \mid $}}#1}       %
\def\Tilde#1{\widetilde{#1}}                    
\def\Hat#1{\widehat{#1}}                        
\def\Bar#1{\overline{#1}}                       
\def\sbar#1{\stackrel{*}{\Bar{#1}}}             
\def\bra#1{\left\langle #1\right|}              
\def\ket#1{\left| #1\right\rangle}              
\def\VEV#1{\left\langle #1\right\rangle}        
\def\abs#1{\left| #1\right|}                    
\def\leftrightarrowfill{$\mathsurround=0pt \mathord\leftarrow \mkern-6mu
        \cleaders\hbox{$\mkern-2mu \mathord- \mkern-2mu$}\hfill
        \mkern-6mu \mathord\rightarrow$}
\def\dvec#1{\vbox{\ialign{##\crcr
        \leftrightarrowfill\crcr\noalign{\kern-1pt\nointerlineskip}
        $\hfil\displaystyle{#1}\hfil$\crcr}}}           
\def\dt#1{{\buildrel {\hbox{\LARGE .}} \over {#1}}}     
\def\dtt#1{{\buildrel \bullet \over {#1}}}              
\def\der#1{{\pa \over \pa {#1}}}                
\def\fder#1{{\d \over \d {#1}}}                 


\def\frac#1#2{{\textstyle{#1\over\vphantom2\smash{\raise.20ex
        \hbox{$\scriptstyle{#2}$}}}}}                   
\def\half{\frac12}                                        
\def\sfrac#1#2{{\vphantom1\smash{\lower.5ex\hbox{\small$#1$}}\over
        \vphantom1\smash{\raise.4ex\hbox{\small$#2$}}}} 
\def\bfrac#1#2{{\vphantom1\smash{\lower.5ex\hbox{$#1$}}\over
        \vphantom1\smash{\raise.3ex\hbox{$#2$}}}}       
\def\afrac#1#2{{\vphantom1\smash{\lower.5ex\hbox{$#1$}}\over#2}}    
\def\partder#1#2{{\partial #1\over\partial #2}}   
\def\parvar#1#2{{\d #1\over \d #2}}               
\def\secder#1#2#3{{\partial^2 #1\over\partial #2 \partial #3}}  
\def\on#1#2{\mathop{\null#2}\limits^{#1}}               
\def\bvec#1{\on\leftarrow{#1}}                  
\def\oover#1{\on\circ{#1}}                              

\def\[{\lfloor{\hskip 0.35pt}\!\!\!\lceil}
\def\]{\rfloor{\hskip 0.35pt}\!\!\!\rceil}
\def\Lag{{\cal L}}
\def\du#1#2{_{#1}{}^{#2}}
\def\ud#1#2{^{#1}{}_{#2}}
\def\dud#1#2#3{_{#1}{}^{#2}{}_{#3}}
\def\udu#1#2#3{^{#1}{}_{#2}{}^{#3}}
\def\calD{{\cal D}}
\def\calM{{\cal M}}

\def\szet{{${\scriptstyle \b}$}}
\def\ulA{{\un A}}
\def\ulM{{\underline M}}
\def\cdm{{\Sc D}_{--}}
\def\cdp{{\Sc D}_{++}}
\def\vTheta{\check\Theta}
\def\fracm#1#2{\hbox{\large{${\frac{{#1}}{{#2}}}$}}}
\def\ha{{\fracmm12}}
\def\tr{{\rm tr}}
\def\Tr{{\rm Tr}}
\def\itrema{$\ddot{\scriptstyle 1}$}
\def\ula{{\underline a}} \def\ulb{{\underline b}} \def\ulc{{\underline c}}
\def\uld{{\underline d}} \def\ule{{\underline e}} \def\ulf{{\underline f}}
\def\ulg{{\underline g}}
\def\items#1{\\ \item{[#1]}}
\def\ul{\underline}
\def\un{\underline}
\def\fracmm#1#2{{{#1}\over{#2}}}
\def\footnotew#1{\footnote{\hsize=6.5in {#1}}}
\def\low#1{{\raise -3pt\hbox{${\hskip 0.75pt}\!_{#1}$}}}

\def\Dot#1{\buildrel{_{_{\hskip 0.01in}\bullet}}\over{#1}}
\def\dt#1{\Dot{#1}}

\def\DDot#1{\buildrel{_{_{\hskip 0.01in}\bullet\bullet}}\over{#1}}
\def\ddt#1{\DDot{#1}}

\def\DDDot#1{\buildrel{_{_{\hskip 0.01in}\bullet\bullet\bullet}}\over{#1}}
\def\dddt#1{\DDDot{#1}}

\def\DDDDot#1{\buildrel{_{_{\hskip 
0.01in}\bullet\bullet\bullet\bullet}}\over{#1}}
\def\ddddt#1{\DDDDot{#1}}

\def\Tilde#1{{\widetilde{#1}}\hskip 0.015in}
\def\Hat#1{\widehat{#1}}


\newskip\humongous \humongous=0pt plus 1000pt minus 1000pt
\def\caja{\mathsurround=0pt}
\def\eqalign#1{\,\vcenter{\openup2\jot \caja
        \ialign{\strut \hfil$\displaystyle{##}$&$
        \displaystyle{{}##}$\hfil\crcr#1\crcr}}\,}
\newif\ifdtup
\def\panorama{\global\dtuptrue \openup2\jot \caja
        \everycr{\noalign{\ifdtup \global\dtupfalse
        \vskip-\lineskiplimit \vskip\normallineskiplimit
        \else \penalty\interdisplaylinepenalty \fi}}}
\def\li#1{\panorama \tabskip=\humongous                         
        \halign to\displaywidth{\hfil$\displaystyle{##}$
        \tabskip=0pt&$\displaystyle{{}##}$\hfil
        \tabskip=\humongous&\llap{$##$}\tabskip=0pt
        \crcr#1\crcr}}
\def\eqalignnotwo#1{\panorama \tabskip=\humongous
        \halign to\displaywidth{\hfil$\displaystyle{##}$
        \tabskip=0pt&$\displaystyle{{}##}$
        \tabskip=0pt&$\displaystyle{{}##}$\hfil
        \tabskip=\humongous&\llap{$##$}\tabskip=0pt
        \crcr#1\crcr}}


\def\eV{\,{\rm eV}}
\def\keV{\,{\rm keV}}
\def\MeV{\,{\rm MeV}}
\def\GeV{\,{\rm GeV}}
\def\TeV{\,{\rm TeV}}
\def\sv{\left<\sigma v\right>}
\def\({\left(}
\def\){\right)}
\def\cm{{\,\rm cm}}
\def\K{{\,\rm K}}
\def\kpc{{\,\rm kpc}}
\def\beq{\begin{equation}}
\def\eeq{\end{equation}}
\def\bea{\begin{eqnarray}}
\def\eea{\end{eqnarray}}


\newcommand{\be}{\begin{equation}}
\newcommand{\ee}{\end{equation}}
\newcommand{\nbe}{\begin{equation*}}
\newcommand{\nee}{\end{equation*}}

\newcommand{\fr}{\frac}
\newcommand{\lb}{\label}

\thispagestyle{empty}

{\hbox to\hsize{
\vbox{\noindent May 2024 \hfill IPMU24-0020} }}

\noindent  

\noindent
\vskip2.0cm
\begin{center}

{\large\bf Einstein-Grisaru-Zanon gravity}

\vglue.4in

Ruben Campos Delgado~${}^{a,\&}$  and Sergei V. Ketov~${}^{b,c,d,\#,}$\footnote{The corresponding author} 
\vglue.3in

${}^a$~Institut f\"ur Theoretische Physik, Leibniz Universit\"at Hannover, \\ Appelstra\ss e 2,
30167 Hannover, Germany\\
${}^b$~Department of Physics, Tokyo Metropolitan University,\\
1-1 Minami-ohsawa, Hachioji-shi, Tokyo 192-0397, Japan \\
${}^c$~Research School of High-Energy Physics, Tomsk Polytechnic University, \\
Tomsk 634028, Russian Federation\\
${}^d$~Kavli Institute for the Physics and Mathematics of the Universe (WPI),
\\The University of Tokyo Institutes for Advanced Study,  Kashiwa 277-8583, Japan\\

\vglue.2in

${}^{\&}$~ruben.camposdelgado@gmail.com, ${}^{\#}$~ketov@tmu.ac.jp
\end{center}

\vglue.4in

\begin{center}
{\Large\bf Abstract}  
\end{center}
\vglue.1in

\noindent The leading $(\alpha')^3$-correction to the gravitational low-energy effective action of  closed (type II) superstring theory in four-spacetime dimensions defines the Einstein-Grisaru-Zanon gravity action that is applied for a calculation of the leading corrections to the Schwarzschild solution and the Hubble function in the Friedmann-Lemaitre-Robertson-Walker universe, in the first order with respect to the effective string-generated coupling. The solutions found are compared to the corresponding solutions in the Einstein-Bel-Robinson gravity that also modifies the Einstein gravity by the terms quartic in the spacetime curvature. We consider the black hole shadows in the Einstein-Grisaru-Zanon gravity theory and derive the upper bound on the string coupling parameter from the Hawking temperature of a black hole.

\newpage

\section{Introduction}

The nonrenormalizability of the pure Einstein gravity in four spacetime dimensions was proved by Goroff and Sagnotti in 1985 by calculating the two-loop (on-shell) ultra-violet (UV) divergence cubic in the spacetime curvature \cite{Goroff:1985sz}.  In the case of $N=1$ supergravity in four spacetime dimensions, the leading  UV-divergence is expected in three loops, being quartic in the space-time curvature. The UV-completion of Einstein gravity in quantum theory is provided by closed (type II) superstring theory, where the 
UV-counterterms of the quantized gravity appear with finite coefficients. The gravitational low-energy effective action (LEEA) in closed superstring theory can be computed
perturbatively, with respect to the string slope parameter $\alpha'$ and the string coupling $g_s$, resulting in the infinite series with respect to the spacetime curvature \cite{Kiritsis:2019npv}. For instance, the perturbative $\alpha'$-corrections are computable by deriving the multi-loop renormalization group beta-functions of the two-dimensional non-linear sigma-model (NLSM) describing propagation of a test string in spacetime, demanding quantum conformal invariance of the NLSM and, hence, the vanishing beta-functions as the effective equations of motion in spacetime, and then elevating those equations of motion to the LEEA in the critical dimension \cite{Fradkin:1985ys}. The existence of such action is assured by Zamolodchikov's $c$-theorem \cite{Zamolodchikov:1986gt}. This procedure reproduces the Einstein-Hilbert action in the one-loop approximation \cite{Callan:1985ia}, whereas the next $\alpha'$-correction derived from the supersymmetric NLSM appears at four loops, being proportional to $(\alpha')^3$ and quartic in the spacetime curvature.~\footnote{See Ref.~\cite{Ketov:1991jw} for the non-supersymmetric gravitational LEEA from bosonic closed string theory and its applications to black holes, with the
 $(\alpha')^2$-correction being cubic in the spacetime curvature.} The leading perturbative correction to the closed superstring gravitational LEEA was first computed by Grisaru and Zanon in 1986 \cite{Grisaru:1986vi}, see also Ref.~\cite{Ketov:2000dy} for more details about its derivation in the NLSM approach. The same quartic curvature terms arise from M-theory \cite{Green:1997as}, see also Ref.~\cite{Blumenhagen:2024ydy} for a recent account.

In this Letter we apply the Einstein-Grisaru-Zanon (EGZ) gravitational LEEA to a derivation of the corresponding corrections to the standard metric solutions in Einstein's General Relativity (GR), and compare them to  the similar solutions in the Einstein-Bel-Robinson (EBR) gravity proposed in Ref.~\cite{Ketov:2022lhx}.~\footnote{The EBR gravity action can be elevated to the Starobinsky-Bel-Robinson (SBR) gravity 
action by adding the $R^2$-term. However, the $R^2$-term does not contribute to the Schwarzshild solution.}

The paper is organized as follows. In Sec.~2 we derive the leading perturbative corrections to the Schwarzschild metric, in the first order
with respect to the effective string theory parameter, both in the EGZ and EBR cases, and compare them. In Sec.~3 we perform a similar
analysis for a spatially flat  Friedmann-Lemaitre-Robertson-Walker (FLRW) universe. In Sec.~4 we apply our results to the observable black hole shadows.  In Sec.~5 we find the upper-bound on the string parameter in the EGZ and EBR actions.  Our conclusion is 
Sec.~6.

\section{EGZ versus EBR}

The EGZ action is defined by
\begin{equation}
    S_{\rm EGZ}[g]=\fracmm{M_{\text{Pl}}^2}{2}\int d^4x \sqrt{-g}\,\left(R+\fracmm{\gamma}{M_{\text{Pl}}^6}J\right), \label{egz}
\end{equation}
where we have introduced the leading superstring correction \cite{Grisaru:1986vi},
\begin{equation}
    J=\left(R^{\mu\rho\sigma\nu}R_{\lambda\rho\sigma\tau}+\frac{1}{2}R^{\mu\nu\rho\sigma}R_{\lambda\tau\rho\sigma}\right)R_{\mu}^{\,\,\,\alpha\beta\lambda}R^{\tau}_{\,\,\,\alpha\beta\nu}~~,
\end{equation}
and the new (dimensionless) coupling constant $\gamma$. The $M_{\rm Pl}$ is the reduced Planck mass, $M_{\rm Pl}=(8\p G_{\rm N})^{-1/2}$. The value of $\gamma$ cannot be calculated from string theory at present, because the action (\ref{egz}) is in four-space dimensions, so that $\gamma$ depends upon a compactification from ten to four dimensions and the unknown vacuum expectation value of string dilaton.
    
The equations of motion from the superstring gravitational LEEA (\ref{egz}) have the structure
\begin{equation}\label{eq:eom1}
    G_{\mu\nu}-\fracmm{\gamma}{M_{\text{Pl}}^6}H_{\mu\nu}=0~,
\end{equation}
where $G_{\mu\nu}$ is Einstein's tensor and $H_{\mu\nu}$ stands for the contribution of $J$.

Let us investigate how  the GZ term affects the Schwarzschild solution in GR, in the first order with respect to $\gamma$,
by using  the standard (spherically-symmetric and static) Ansatz for the metric,
\begin{equation} \lb{man}
    ds^2=-A(r)dt^2+\fracmm{1}{B(r)}dr^2+r^2d\Omega^2_2~,
\end{equation}
with two functions $A(r)$ and $B(r)$, having the form

\begin{equation}
    A(r)=1-\fracmm{2G_{\rm N}M}{r}-\gamma \Delta(r), \hspace{6mm}  B(r)=1-\fracmm{2G_{\rm N}M}{r}-\gamma \Sigma(r)~,
\end{equation}
in terms of two unknown functions $\D(r)$ and $\S(r)$, outside a spherically-symmetric body of mass $M$.

For this purpose, we have evaluate $H_{\mu\nu}$ only on the classical Schwarzschild solution, because  $H_{\mu\nu}$ is multiplied by 
$\gamma$ in the equations of motion (\ref{eq:eom1}).  Using the {\it xTras} package of Mathematica \cite{Nutma:2013zea} we derived the  following solution:
\begin{equation}\label{eq:A}
A(r)=  1-\fracmm{2G_{\rm N}M}{r}-\gamma \fracmm{2^{11}\pi^3 G^6_{\rm N} M^3(8r-11G_{\rm N}M)}{r^{10}}
\end{equation}
and
\begin{equation}\label{eq:B}
B(r)=  1-\fracmm{2G_{\rm N}M}{r}-\gamma \fracmm{2^{11}\pi^3 G^6_{\rm N} M^3(36r-67G_{\rm N}M)}{r^{10}}~~.
\end{equation}

It is instructive to compare these results with the EBR gravity arising from the SBR gravity \cite{Ketov:2022lhx} in the absence of  the $R^2$-term and having the action with the terms quartic in the curvature  also,
\be
S_{\rm EBR} [g] =  \fracmm{M_{\rm Pl}^2}{2}\int d^{4}x\,\sqrt{-g}\left( R -\fracmm{\b}{32M_{\rm Pl}^6} T^2 \right)~,
\label{ebr}
\ee
where $\b$  is another dimensionless coupling constant, the $T^2$ stands for the Bel-Robinson (BR) tensor squared,
$T^2\equiv T_{\r\n\l\m}T ^{\r\n\l\m}$, with \cite{Bel:1959uwe,Robinson:1997}
\be
T^{\r \n\l\m} \equiv R^{\r \s \h\l}R\udu{\n}{\s \h}{\m} +{}^*R^{\r \s \h\l}{}^*R\udu{\n}{\s \h}{\m} =R^{\r \s \h\l}R\udu{\n}{\s\h}{\m}
 +R^{\r \s \h\m}R\udu{\n}{\s \h}{\l}-\ha g^{\r\n}R^{\s \h \x\l} R\du{\s \h \x}{\m}~,
\label{belr}
\ee
and the star denotes the Hodge dual tensor in four dimensions. The BR tensor was introduced by Bel and Robinson \cite{Bel:1959uwe,Robinson:1997} by analogy with the standard energy-momentum tensor of Maxwell's theory of electromagnetism,
\be T^{\rm Maxwell}_{\m\n}=F_{\m\r}F\du{\n}{\r}+{}^*F_{\m\r}{}^*F\du{\n}{\r}~~,\quad 
F_{\m\n}=\pa_{\m}A_{\n}-\pa_{\n}A_{\m}\lb{max}~~.
\ee

The BR tensor squared obeys the identity \cite{Deser:1999jw,Iihoshi:2007vv}
\be
T^2=  -\frac{1}{4}({}^*\!R_{\m\n\l\r}{}^*\!R^{\m\n\l\r})^2 
+\frac{1}{4}({}^*R_{\m\n\l\r}R^{\m\n\l\r})^2 
 = \frac{1}{4}(P_4^2-E^2_4) =\frac{1}{4}(P_4+E_4)(P_4-E_4)~, \label{id2}
\ee
where the Euler and Pontryagin topological densities in four dimensions, $E_4$ and $P_4$, respectively, have been introduced. It is worth recalling yet another well-known identity, 
\be \lb{gb}
E_4 = {\cal G}_{\rm GB} \equiv R^{\m\n\l\r}R_{\m\n\l\r} - 4R^{\m\n}R_{\m\n} + R^2~,
\ee
where the Gauss-Bonnet invariant ${\cal G}_{\rm GB}$ has been introduced, because it gives the connection between the
EBR gravity theory (\ref{ebr}) and the modified gravity theories with the Lagrangians of the form $\Lag(R,{\cal G}_{\rm GB})$. In particular, 
this connection is useful to claim $\beta>0$ \cite{DeFelice:2009ak}.

We consider the EGZ and EBR gravity theories merely perturbatively, in the first order with respect to the new (string) coupling. Therefore, the issue of ghosts does not arise because there are no new degrees of freedom. The BR term was motivated in Ref.~\cite{Ketov:2022zhp} as another "string-inspired" quantum correction, possibly originating from a non-trivial scalar potential of dilaton and axion in the limit of ignoring their kinetic terms (applicable for  slow-roll inflation). The difference between the EGZ and EBR gravity actions in the context 
of superstrings/M-theory was noticed in Ref.~\cite{Moura:2007ks}, see also Ref.~\cite{Chen:2021qrz} for the EGZ gravity in higher dimensions.

When using the same Ansatz for a solution to the EBR theory as in Eq.~(\ref{man}),
 \begin{equation}
    ds^2=-C(r)dt^2+\fracmm{1}{E(r)}dr^2+r^2d\Omega^2_2~,
\end{equation}
with
\begin{equation}
    C(r)=1-\fracmm{2G_{\rm N}M}{r}-\beta \X (r)~, \hspace{6mm}  E(r)=1-\fracmm{2G_{\rm N}M}{r}-\beta\Phi (r)~,
\end{equation}
we find
\begin{equation}
C(r)=  1-\fracmm{2G_{\rm N}M}{r}-\beta \fracmm{2^{11}\pi^3 G^6_{\rm N} M^3(8r-11G_{\rm N}M)}{r^{10}}
\end{equation}
and
\begin{equation}
E(r)=  1-\fracmm{2G_{\rm N}M}{r}-\beta \fracmm{2^{11}\pi^3 G^6_{\rm N} M^3(36r-67G_{\rm N}M)}{r^{10}}~,
\end{equation}
which is the {\it same\/} solution as that in Eqs.~(\ref{eq:A}) and (\ref{eq:B}) for EGZ gravity after the identification of
$\gamma$ and $\beta$. This solution can also be recovered from an independent calculation in Ref.~\cite{Sajadi:2023bwe} in the
case of the vanishing cosmological constant. The black hole thermodynamic quantities in EBR gravity, based on the solution above
were also computed in Refs.~\cite{Sajadi:2023bwe,CamposDelgado:2022sgc}. Therefore, those results are also valid in the EGZ gravity.

Quantum gravity corrections in the gravitational LEEA imply a deviation from the Einstein gravity relation $A(r)=B(r)$ or $C(r)=E(r)$. After rewriting $A(r)=1-2\tilde{A}(r)$ and $B(r)=1+2\tilde{B}(r)$, the post-Newtonian function is defined by \cite{DeFelice:2010aj}
 \be\lb{postN}
\gamma_{\text{pN}}(r)=\fracmm{\tilde{A}(r)}{\tilde{B}(r)}~.
\ee 
 
For a spherically symmetric body with mass $M_c$ and radius $r_c$, the post-Newtonian parameter $\g_{\rm pN}(r_c)$  in the leading order with respect to
the string parameter $\gamma$ in the EGZ and EBR gravity is thus given by
\begin{equation} \lb{pNp}
    \gamma_{\text{pN}}=1+\fracmm{7\cdot 2^{12}\pi^3G^5_{\rm N}M^2_c(2G_{\rm N}M_c-r_c)\gamma}{r^9_c}~~.
\end{equation}

The local gravity constraints on the post-Newtonian parameter inside the Solar system were obtained by Cassini satellite mission 
with the upper bound \cite{Bertotti:2003rm}
\begin{equation} \lb{pNS}
|\gamma_{\text{pN}}-1|<2.3\times 10^{-5}~.
\end{equation}
However, this bound does not lead to a meaningful constraint on the parameter $\gamma$ when using Eq.~(\ref{pNp}) in application to the Sun or black holes, either primordial or supermassive ones.

\section{EGZ and EBR in FLRW}

The difference between the EGZ and EBR gravity becomes apparent in a spatially flat FLRW universe characterized by the metric
\begin{equation}
    ds^2=-dt^2+a^2(dx_1^2+dx_2^2+dx_3^2)~,\end{equation}
where $a(t)$ is the cosmic scale factor. 

In both cases (EBR and EGZ) we derived two equations of motion, corresponding to the $tt$ and $x_ix_i$ (no sum) components, and verified their consistency. 

In terms of the (observable)  Hubble function $H=\dot{a}/a$, the equations of motion in the EBR theory are given by 
\begin{equation}\label{eq:EBRtt}
    H^2+\fracmm{3\beta}{M_{\text{Pl}}^6}H^4\left(-H^4+2H\ddot{H}+6H^2\dot{H}+3\dot{H}^2\right)=0
\end{equation}
and
\begin{equation}\label{eq:EBRxx}
\begin{gathered}
    3H^2+2\dot{H}-\fracmm{3\beta}{M_{\text{Pl}}^6}H^2\Big[3H^6-12H^3\ddot{H}-16H\dot{H}\ddot{H}-12\dot{H}^3\\
    +2H^4\left(-6\ddot{H}+\dot{H}\right)-H^2\left(2H^{(3)}+45\dot{H}^2\right)\Big]=0.
\end{gathered}
\end{equation}
In particular, Eq.~\eqref{eq:EBRtt} is consistent with the analogous equation derived in Ref.~\cite{Ketov:2022zhp}.

In the EGZ gravity, the equations of motion are
\begin{equation}\label{eq:EGZtt}
\begin{gathered}
    H^2+\fracmm{\gamma}{M_{\text{Pl}}^6}\Big[H^4\left(-12H^4+44H\ddot{H}+132H^2\dot{H}+138\dot{H}^2\right)\\
    +48H^3\dot{H}\ddot{H}+12H\dot{H}^2\ddot{H}+28H^2\dot{H}^3-3\dot{H}^4\Big]=0
\end{gathered}
\end{equation}
and
\begin{equation}
\begin{gathered}
3H^2+2\dot{H}+\fracmm{\gamma}{M_{\text{Pl}}^6}H^2\Big\{4H^2\left(-9H^6+66H^3\ddot{H}+66H^4\ddot{H}+12\ddot{H}^2+11H^2H^{(3)}\right)\\
+4\dot{H}\left(9H^6+160H^3\ddot{H}+72H^4\ddot{H}+6\ddot{H}^2+12H^2H^{(3)}\right)\\
+6\dot{H}^2\left(153H^4+4H\ddot{H}(11+3H)+2H^{(3)}\right)+564H^2\dot{H}^3+47\dot{H}^4\Big\}=0~
\end{gathered}
\end{equation}
where the $H^{(3)}$ is the third time derivative. A simple comparison of the equations of motion in the EBR and EGZ theories shows they are different, except the trivial case when all the time derivatives of the Hubble function are ignored.

The reason why the solutions to the EGZ and EBR gravity theories are the same in the Schwarzschild case can be explained by the observation 
that in the case of $R=0$ and $R_{\mu\nu}=0$, one also has $P_4=0$ and 
\begin{equation}
    J-\fracmm{\mathcal{G}^2}{32}=\fracmm{1}{8}R_{\alpha\beta}^{\,\,\,\,\,\,\,\rho\sigma}R^{\alpha\beta\gamma\delta}R_{\gamma\rho}^{\,\,\,\,\,\,\,\lambda\tau}R_{\delta\sigma\lambda\tau}-\fracmm{1}{16}R_{\alpha\beta}^{\,\,\,\,\,\,\,\rho\sigma}R^{\alpha\beta\gamma\delta}R_{\gamma\delta}^{\,\,\,\,\,\,\,\lambda\tau}R_{\rho\sigma\lambda\tau}=0~,
\end{equation}
whereas both the Ricci scalar and the Ricci tensor do not vanish for the FLRW metric. 

The solutions to the EBR gravity theory in the FLRW universe were obtained in Ref.~\cite{Ketov:2022zhp} in the more general case including
the $R^2$-term (dubbed the SBR gravity). To the end of this Section, we focus on the EGZ gravity. It admits the constant (de Sitter) solution with
\begin{equation}\label{eq:const}
    H_0=\fracmm{M_{\text{Pl}}}{(12\gamma)^{1/6}}>0.
\end{equation}

Let us consider small perturbations around $H_0$ with $H(t)=H_0+\delta H (t)$. The linearized version of Eq.~\eqref{eq:EGZtt}
is then given by
\begin{equation}
-\fracmm{72M_{\text{Pl}}\gamma}{(12\gamma)^{7/6}}\delta H(t)+11\delta \dot{H}(t)+\fracmm{11(12\gamma)^{1/6}}{3M_{\text{Pl}}}\delta \ddot{H}(t)=0~,
\end{equation}
whose solution reads
\begin{equation}
    \delta H(t)=c_{+}e^{\alpha_+t}+c_{-}e^{\alpha_{-}t}~,
\end{equation}
where $c_{\pm}$ are the integration constants and
\begin{equation}
    \alpha_{\pm}=\pm\fracmm{(\mp 33 + 3\sqrt{209})M_{\text{Pl}}}{22(12\gamma)^{1/6}}~~.
\end{equation}
Since $\alpha_+$ is always positive, the de Sitter solution \eqref{eq:const} is unstable.

A more general  time-dependent solution can be found in the form of Taylor (or Painleve) expansion,
\begin{equation}
    H(t)=\sum_{k=-p}^{+\infty}c_k(t-t_0)^k~,
\end{equation}
where $t_0$ is an integration constant and $p$ is a positive integer. We find in the EGZ gravity that the solution to Eq.~\eqref{eq:EGZtt} does not include the negative powers, while the first terms of the expansion are given by
\begin{equation}\label{eq:sol_friedmann}
    H(t)=\fracmm{M_{\text{Pl}}}{(12\gamma)^{1/6}}+\fracmm{13}{21}\left(\fracmm{2}{3\gamma}\right)^{1/3}M_{\text{Pl}}^2(t-t_0)-\fracmm{3445M_{\text{Pl}}^3}{1617\sqrt{3\gamma}}(t-t_0)^2+\mathcal{O}\left((t-t_0)^3\right).
\end{equation}
In terms of the dimensionless quantities
\begin{equation}
    \tilde{H}\equiv\fracmm{H_0}{M_{\text{Pl}}} \quad {\rm and} \quad  \lambda(t)\equiv M_{\text{Pl}}(t-t_0)~,
\end{equation}
the solution \eqref{eq:sol_friedmann} can be rewritten to
\begin{equation}
    \fracmm{H(t)}{M_{\text{Pl}}}=\tilde{H}\left[1+\fracmm{13}{21}\left(\fracmm{16}{3\gamma}\right)^{1/6}\lambda(t)-\fracmm{3445}{1617}\left(\fracmm{4}{3\gamma}\right)^{1/6}\lambda^2(t)\right] + \mathcal{O}(\left(\l^3(t)\right)~.
\end{equation}

\section{Black hole shadows}

In this Section we briefly revisit the earlier studies of black hole shadows and confront them with observations in the context of the
superstring gravity, see e.g.~Refs.~\cite{Arora:2023ijd,Belhaj:2023dsn,Bolokhov:2023dxq,Agurto-Sepulveda:2024iwu}, 
by using our results in Eqs.~\eqref{eq:A} and \eqref{eq:B}. 

The angular radius of a black hole shadow is given by \cite{Papnoi:2021rvw,Perlick:2021aok}
\begin{equation}
    \sin^2\alpha_{\rm sh}=\fracmm{h(r_{\rm ph})^2}{h(r_{\rm obs})^2}~~,
\end{equation}
where
\begin{equation}
    h(r)=\fracmm{g_{\theta\theta}}{g_{rr}}=\fracmm{r^2}{A(r)}~~,
\end{equation}
and $r_{\rm ph}$, $r_{\rm obs}$ are the photon sphere radius and the observer distance, respectively. For an observer far away from a black hole, the observable radius of the black hole shadow is 
\begin{equation}\label{eq:shadow}
    R_{\rm sh}=r_{\rm obs}\alpha_{\rm sh}\approx \fracmm{r_{\rm ph}}{\sqrt{A(r_{\rm ph})}}~~.
\end{equation}

The photon sphere radius can be obtained from the condition \cite{Papnoi:2021rvw,Perlick:2021aok}
\begin{equation}
    \fracmm{d}{dr}h^2(r)\Big|_{r=r_{\rm ph}}=0~.
\end{equation}
In the first order with respect to $\gamma$ or $\beta$, we find 
\begin{equation}
    r_{\rm ph}=3G_NM+\gamma \fracmm{45056\pi^3}{6561G^2_{\rm N}M^5}~~,
\end{equation}
where $M$ is the black hole mass. Plugging this result into Eq.~\eqref{eq:shadow} gives
\begin{equation} \lb{shR}
    R_{\rm sh}=3\sqrt{3}G_{\rm N}M+\gamma\fracmm{13312\pi^3}{2187\sqrt{3}G^2_{\rm N}M^5}~~.
\end{equation}

Astrophysical observations with the Event Horizon Telescope measured the angle diameter of the shadow of the black hole $M87^{\ast}$ in the center of the galaxy M87 as $\Omega= 42\pm3$ $\mu$ \cite{EventHorizonTelescope:2019dse}. The distance between the black hole and the Solar system is estimated as $D = 16.8\pm0.8$ Mpc. This gives the diameter of the black hole shadow as $d_{\rm sh} = 2R_{\rm sh} = D\Omega$ and allows us to derive the following expression for $\gamma$:
\begin{equation}\label{eq:gamma_shadow}
    \gamma=\fracmm{2187\sqrt{3}G^2_{\rm N}M^5}{26624\pi^3\hbar^3c}\left(d_{\rm sh}-6\sqrt{3}G_{\rm N}M/c^2\right).
\end{equation}
When using the known distance $9.33497 \cdot 10^{13}~{\rm m} < d_{\rm sh} < 1.1552 \cdot 10^{14}$ m and the known mass of $M87^{\ast}$ black hole as 
$M = (6.5 \pm 0.7) \cdot 10^9 M_{\odot}$,  one might derive the upper bound on $\gamma$ from this equation, similarly to
Ref.~\cite{Arora:2023ijd}. However, due to the uncertainties in measuring both $d_{\rm sh}$ and $M$, it requires an enormous fine-tuning between the terms in the curved brackets of Eq.~(\ref{eq:gamma_shadow}) where we have  $6\sqrt{3}G_{\rm N}M/c^2\approx 9.98 \cdot 10^{13}$ m and the multiplying factor $\fracmm{2187\sqrt{3}G^2_{\rm N}M^5}{26624\pi^3\hbar^3c}\approx 8.47\times 10^{268}$ m$^{-1}$.
Therefore, the claim of Ref.~\cite{Arora:2023ijd} about the upper bound on $\gamma$ of the order $10^{-3}$ should be dismissed.
 
\section{The upper bound on $\gamma$ from Hawking temperature}
 
A better way to derive the upper bound comes from the Hawking temperature of a black hole. In general, a quantum gravity correction should be
small enough for perturbation theory to make sense. The radius of the event horizon is obtained as a solution to 
\begin{equation}
A(r)=0 \quad {\rm or }\quad B(r)=0 ~~.
\end{equation}
In the first order with respect to $\gamma$, it implies
\begin{equation}
    r_{\rm h}=2G_{\rm N}M+\gamma\fracmm{20\pi^3}{G^2_{\rm N}M^5}~~.
\end{equation}

The Hawking temperature is given by
\begin{equation}
    T=\fracmm{\sqrt{A'(r_{\rm h})B'(r_{\rm h})}}{4\pi}~~,
\end{equation}
where the primes denote the derivatives. After reinstating the fundamental constants, we find
\begin{equation}
    T=\fracmm{\hbar c^3}{8\pi G_{\rm N}M k_B}\left(1-\gamma \fracmm{8\pi^3\hbar^3c^3}{G_{\rm N}^3M^6}\right)~~,
\end{equation}
that is in agreement with the well-known statement that the Hawking temperature of a black hole is decreased in superstring gravity
\cite{Myers:1987qx}. Therefore, we have
\begin{equation}\label{eq:constraint}
    \gamma<\fracmm{G_{\rm N}^3M^6}{8\pi^3\hbar^3c^3}
\end{equation}
for {\it any\/} black hole of mass $M$. Taking the smallest black hole with the Planck mass $M_{\rm Pl}\approx 2.176\cdot 10^{-8}$ kg
in \eqref{eq:constraint} yields
\begin{equation} \lb{gammaup}
    \gamma < 1.62 \cdot 10^{-5}~~.
\end{equation}
This bound is stronger than the bounds found in Refs.~\cite{Belhaj:2023dsn, Bolokhov:2023dxq}.

\section{Conclusion}

Our main results are given by Eqs.~(\ref{eq:A}),  (\ref{eq:B}),  (\ref{pNp}), (\ref{shR}) and  (\ref{gammaup}).

Equations (\ref{eq:A}) and (\ref{eq:B}) give the perturbative solution to the EGZ gravity
in the first order with respect to the string coupling parameter $\gamma$, which modifies the standard Schwarzschild solution by the leading closed superstring correction in the gravitational LEEA.

Equation (\ref{pNp}) gives the post-Newtonian parameter in the EGZ gravity for a spherically symmetric and static source of mass $M_c$ and radius $r_c$.

Equation (\ref{shR}) gives the EGZ correction to the radius of the black hole shadow, and
Eq.~(\ref{gammaup}) gives the upper bound on the value of $\gamma$.

\section*{Acknowledgements}

The authors are grateful to Filipe Moura and Julio Oliva Zapata for correspondence.

SVK was supported by Tokyo Metropolitan University, the Japanese Society for Promotion of Science under the grant No.~22K03624, the World Premier International Research Center Initiative (MEXT, Japan), and the Tomsk Polytechnic University development program 
Priority-2030-NIP/EB-004-375-2023.

\bibliographystyle{utphys}
\bibliography{references}

\end{document}